# Flat quasicrystalline tilings in terms of density wave approach

Aleksey S. Roshal[1], Olga V. Konevtsova[1] and Sergei B. Rochal[1,*]
[1]*Faculty of Physics, Southern Federal University, 344090 Rostov-on-Don, Russia*

Classical Landau theory considers structural phase transitions and crystallization as a condensation of several critical density waves whose wave vectors are symmetrically equivalent. Analyzing the simplest nonequilibrium Landau potentials obtained for decagonal and dodecagonal cases, we derive constraints on the phases of the critical waves and deduce two pairs of flat tilings that are the simplest from the viewpoint of our theory. Each pair corresponds to the same irreducible interference pattern: the vertices of the first and second tilings are located at its minima and maxima, respectively. The first decagonal pair consists of the Penrose P1 tiling and the Tie and Navette one. The second pair is represented by dodecagonal tiling of squares, triangles, and shields, and previously unidentified one formed by regular dodecagons and identical deformed pentagons. Surprisingly, the proposed method for finding extrema of interference patterns provides a straightforward way to generate the Penrose tiling P3 and its more complicated analogues with 2n-fold symmetries. Within Landau theory, we discuss the assembly of the square-triangular tiling and its relationship with the dodecagonal tiling that includes shields. Then we develop a nonequilibrium assembly approach that is based on Landau theory and allows us to produce tilings with random phason strain characteristic of quasicrystals. Interestingly, the approach can generate tilings without or with a minimum number of defective tiles. Examples of real systems rationalized within Landau theory are considered as well. Finally, the derivation of other tilings arising from the reducible interference patterns is discussed, and the relative complexity of non-phenomenological interactions required for the assembly of decagonal and dodecagonal structures is analyzed.

## I. INTRODUCTION.

The first quasicrystal (QC) discovered in Al-Mn alloy about 40 years ago [1] had icosahedral symmetry, which is impossible in periodic structures. Over the next decade, it was found that various icosahedral, octa-, and decagonal QCs are quite common for binary and ternary metal compounds [2], in which QCs appear to be thermodynamically stable states at temperatures slightly below the melting point of the system. A characteristic feature of QC structures is the unique aperiodic long-range order, which makes Bragg diffraction possible. Subsequently, dodecagonal QCs were revealed also in a number of soft-matter systems, among which one can highlight dendritic liquid quasicrystals formed from complex treelike molecules [3], colloidal nanocrystals self-assembling in an aqueous solution from micelles with core-shell structure [4], various polymer QCs [5,6], and other similar systems [7-8]. In recent years, dodecagonal order was discovered in metal-oxide layers [9-16], binary nanocrystal superlattices [4,17-18], bilayer graphene [19-21], and, locally, even in some protein shells [22-23].

Various methodologies exist for the structural description of QCs [24-26], but almost all of them trace back to the ideas of n-dimensional (nD) crystallography, which posits that incommensurate aperiodic structure with long-range order can be considered as a periodic one in a higher dimensional space [27]. In particular, using nD crystallography one can construct aperiodic tilings formed by several types of identical tiles, and then decorate these tiles with a certain molecular or atomic structure [24].

After the discovery of QCs, Landau crystallization theory and similar approaches were used to explain their structure and properties [28-30]. In Landau theory, the energy of the emerging low-symmetry phase is determined by amplitudes of static critical density waves [31-32], whose wave vectors can form stars with non-crystallographic symmetry and the corresponding wave interference patterns can be aperiodic. The critical density waves span a single irreducible representation of a highly symmetric phase (melt or solution). Due to this, their wave vectors are symmetrically equivalent and formally they can be considered as projections of equivalent vectors from the reciprocal space of the corresponding nD periodic lattice. Thus, the interference patterns obtained within the framework of Landau theory can be related with tilings known from nD crystallography. However, these relations can be obtained for a few tilings only, and since Landau theory is rooted on fundamental physical principles, structures based on such tilings should be the most common in nature.

In this work, we consider the tilings that result from Landau theory in the decagonal and dodecagonal cases. As we show within this theoretical framework, tilings with the same rotational symmetry arise in dual pairs: the vertices of each tiling in the pair correspond to the positions of either the most pronounced maxima or minima of the irreducible interference pattern. Analyzing the possible interference patterns, we find, in particular, a dodecagonal tiling that has been observed in nature but its structure has not been comprehensively explained before.

Using critical density waves it is possible to construct pair potentials [28-32] which, when applied in the nonequilibrium assembly model, produce random tilings composed of the same tile types as regular tilings. This approach implemented for octagonal Ammann-Beenker (AB) tiling was reported in Ref. [33]. In this paper, we develop a

*Contact author: rochal_s@yahoo.fr

nonequilibrium assembly model for the pentagonal Penrose tiling P1, composed of pentagons, (truncated) stars, and narrow rhombuses [34-35]. In the model, a particle can attach to the growing cluster only in those positions, which are separated from the nearest particles by at least one translation from a set of orientationally equivalent ones. The choice of the attachment position is based on the Boltzmann distribution function that depends on the ratio between binding energies calculated for different attachment options and thermal energy. When using suitable model parameters, the resulting random P1 tiling contains no defective tiles, exhibits an average 10-fold symmetry, and demonstrates a high degree of long-range order.

In the following Sections II and III, within the framework of Landau theory we derive two pairs of dual decagonal and dodecagonal tilings and discuss real structures similar to them. In particular, in Section III, we deduce the principles of structural organization of columnar liquid crystal [36] and its periodic approximants. We also consider how to obtain a square-triangular (S-T) tiling in the framework of Landau theory. Further, in Section IV, we apply the approach of critical density waves to construct pair potentials and to model nonequilibrium assembly. In the last Section, we discuss the place of the proposed theory, its limitations and complexity level of the interactions required to assemble the considered QC tilings. Since the tilings obtained within the density wave approach turn out to be specific sublattices of P3-type tilings, in Appendix A we consider a simple method for generating such tiling with arbitrary rotational symmetry. In Appendix B we attempt to assemble Penrose P3 tiling within the model of nonequilibrium assembly.

## II. LANDAU CRYSTALLIZATION THEORY FOR THE DECAGONAL TILINGS

In nature, among the flat QCs, octagonal, decagonal, and dodecagonal tilings are the most common. Landau crystallization theory for the octagonal case was discussed earlier [33], so in this Section and the following ones, we will consider decagonal and dodecagonal cases. Recall that within Landau theory, near the phase transition point, the density distribution of structural units (SUs) is represented as:

$$\rho = \rho_0 + \delta\rho, \qquad (1)$$

where $\rho_0$ is the average density of the initial highly symmetric phase, $\delta\rho(\mathbf{R})$ is a specific perturbation of the initial density describing the SU ordering, and $\mathbf{R}$ is the radius vector. Crystallization causes spontaneous symmetry breaking: in the low-symmetry ordered phase, the addition $\delta\rho \neq 0$ appears, the main contribution to which is made by plane static density waves that span one irreducible representation [33-34].The rotational symmetry of the emerging structure is considered to be given, and in the decagonal case, the star of symmetry-equivalent wave vectors $\mathbf{b}_i^{||}$ of these waves includes 10 rays.

Following [37] we write the plane-wave expansion of $\delta\rho$ as:

$$\delta\rho(\mathbf{R}) = \sum_{k=0}^{9} \rho_k \exp\left(2\pi I \mathbf{b}_k^{||}\mathbf{R}\right), \qquad (2)$$

where $\mathbf{R}$ is a radius vector; $I = \sqrt{-1}$; if $k$=0,1..4 then $\mathbf{b}_k^{||} = b_0 \langle \cos\frac{2k\pi}{5}, \sin\frac{2k\pi}{5} \rangle$ else $\mathbf{b}_{k+5}^{||} = -\mathbf{b}_k^{||}$; $b_0$ is a length of the wave vectors; $\rho_k = |\rho_k| \exp(2\pi I \phi_k)$ are the wave complex amplitudes. Since the density deviation $\delta\rho(\mathbf{R})$ is real, $\rho_k = \rho_j^*$ for $j$=($k$+5) mod 10. In the explicitly real form, the number of independent phases $\phi_k$ is equal to 5, and $\delta\rho$ reads

$$\delta\rho(\mathbf{R}) = \sum_{k=0}^{4} A_k \cos\left(2\pi\left(\mathbf{b}_k^{||}\mathbf{R} + \phi_k\right)\right), \qquad (3)$$

where $A_k = 2|\rho_k|$.

Nonetheless, initially it is more convenient to write the Landau nonequilibrium potential as a function of complex amplitudes:

$$F = F_1 I_1 + F_2 I_2 + F_3 I_1^2 + F_4 I_3 + F_5 I_2 I_1 + F_6 I_1^3, \qquad (4)$$

where $I_1 = \rho_0\rho_5 + \rho_1\rho_6 + \rho_2\rho_7 + \rho_3\rho_8 + \rho_4\rho_9$, $I_2 = \rho_0\rho_1\rho_5\rho_6 + \rho_1\rho_2\rho_6\rho_7 + \rho_2\rho_3\rho_7\rho_8 + \rho_3\rho_4\rho_8\rho_9 + \rho_4\rho_5\rho_9\rho_0$, $I_3 = \rho_0\rho_2\rho_4\rho_6\rho_8 + \rho_1\rho_3\rho_5\rho_7\rho_9$. The value of $F_1$ is a phenomenological coefficient dependent on environmental parameters, and the contribution of the last two six-degree terms ensures the global minimality of the potential. The invariants $I_i$ are unchanged under the permutations of amplitudes $\rho_k$ by the symmetry group of the star $C_{10v}$. In addition, the invariants do not depend on translations, therefore each term in the invariants corresponds to a closed polygonal chain with edges $\pm\mathbf{b}_i^{||}$. If vectors $+\mathbf{b}_i^{||}$ and $-\mathbf{b}_i^{||}$ appear a different number of times in the chain, then the invariant depends on the phases $\phi_k$ in real form. In the decagonal case, such a dependence appears in the 5[th] degree invariant $I_3 \sim \cos(2\pi c_0)$, where $c_0 = \phi_0 + \phi_1 + \phi_2 + \phi_3 + \phi_4$. So, we limited the series (4) to the 6[th] degree.

One can parameterize the phases as $\phi_k = \mathbf{b}_k^{||}\mathbf{r}_0 + \mathbf{b}_k^{\perp}\mathbf{r}_0^{\perp} + \Delta\phi$, where $\mathbf{b}_k^{\perp} = b_0 \langle \cos\frac{6k\pi}{5}, \sin\frac{6k\pi}{5} \rangle$. It is useful since the values $\mathbf{r}_0$, $\mathbf{r}_0^{\perp}$, and $\Delta\phi$ belong to different orthogonal spaces, and only a change in the vector $\mathbf{r}_0$ shifts the interference pattern as a whole. In addition, Landau potential does not depend on the vector $\mathbf{r}_0^{\perp}$ (as well as on the vector $\mathbf{r}_0$), since the closedness of the polygonal chain formed by the

vectors $\mathbf{b}_k^{||}$ leads to the closedness of the polygonal chain formed by the vectors $\mathbf{b}_k^{\perp}$.

Since in the quasicrystalline phase all wave amplitude modules are equal, the resulting effective potential turns out to be a function of two variables: $\Delta\phi$ and $\rho_{av} = |\rho_k|$. Minimizing the potential with respect to $\Delta\phi$ leads to the above condition on $c_0$. Precisely, if $F_4 < 0$ (the coefficient before the 5th degree invariant is negative), the phase combination $c_0$ must be integer, and if the coefficient is positive, $c_0$ must be half integer.

Note that in the standard version of Landau theory, the components of the order parameter (OP) are independent variables. However, if the OP is associated with atomic displacements and the appearance of certain superstructure reflections, then the OP components are proportional to the structural amplitudes of these reflections. When the atomic displacements and OP components are small, there is no point in considering the explicit dependence of the OP components on the atomic coordinates. In contrast, in the present version of Landau theory the amplitudes $\rho_k$ are not small. So, we take into account the dependence of their values on the SU coordinates, which allows us to find the latter. In the QC state the amplitude $\rho_k$ of critical plane waves must be equal, the phases $\phi_k$ are fixed by the constraints derived from potential minimization, and the effective OP $\rho_{av}$ must have the symmetry of the QC state. Thus, we express $\rho_{av}$ as an averaged value of $\rho_k$:

$$\rho_{av} = \frac{1}{5n}\sum_{k=0}^{4}\sum_{j=1}^{n}\cos\left(2\pi\left(\mathbf{b}_k^{||}\mathbf{r}_j + \phi_k\right)\right), \qquad (5)$$

where $n$ is the number of SUs in the considered area of the tiling and $\mathbf{r}_j$ are their coordinates. A comparison of Eq. (5) and Eq. (3) demonstrates that the value of $\rho_{av}$ reaches its maximum when the SUs are located at the maxima of the function $\delta\rho(\mathbf{R})$ (it is taken with equal amplitudes $A_k$). Such a model assumes that the QC is already far enough from the crystallization point, but the main contribution to $\delta\rho$ is still made by functions that span a single irreducible representation. Accordingly, the minimum of the Landau potential is ensured not by its specific dependence on the OP, but by the fact of OP saturation. Analogous variant of Landau theory, in which the OP depends on the coordinates of atoms, and the Landau potential has minima associated with the extremality of the latter dependence, was considered in Ref. [38]. Let us emphasize that placing the SUs at the maxima of $\delta\rho$ function seems entirely natural to us, and the reasoning in this paragraph serves to support this assumption.

Since there are two possible choices for the $F_4$ sign, the developed theory yields two significantly different tilings. In Ref. [37], only the case of $F_4 < 0$ was partially considered: it was shown that if $c_0 = 1$ then the vertices of the Penrose tiling P1 coincide with the most intense maxima of the corresponding interference pattern. However, by superimposing the function $\delta\rho$ on the tiling P1, one can check that the most intense maxima of the functions with different integer $c_0$ always correspond to the P1 tiling.

Unlike traditional nD crystallography, in this work, we obtain tilings by finding the positions of the most intense maxima of $\delta\rho(\mathbf{R})$. Such maxima are located in close proximity to the points $\mathbf{r}_j$, for which the values $\zeta_k = round\left(\mathbf{b}_k^{||}\mathbf{r}_j + \phi_k\right) - \mathbf{b}_k^{||}\mathbf{r}_j - \phi_k$ are concurrently small. It is more convenient to use the single small parameter $\Delta = \sqrt{\Sigma_k\,\zeta_k^2}$, which we denote as the total mismatch. If this parameter is small at the point $\mathbf{r}_j$, then the function $\delta\rho\left(\mathbf{r}_j\right)$ is close to its maximum value.

We obtain the coordinates $\mathbf{r}_j$ from a set of closely located points on the plane $\mathbf{R}$. Firstly, for each starting point $\mathbf{R}^j$ we obtain 5D integer indices as $n_k^j = round\left(\mathbf{b}_k^{||}\mathbf{R}^j + \phi_k\right)$, where $j$=0,1..4, and after that we return to the $\mathbf{R}$ space by applying the expression

$$(\mathbf{R}^j)' = \sum_{k=0}^{4} n_k^j \mathbf{a}_k^{||}, \qquad (6)$$

where $\mathbf{a}_k^{||} = \frac{5}{2b_0}\mathbf{b}_k^{||}$. When projecting into 5D space and back, points that were initially close may end up coinciding. Namely, one can tessellate the plane $\mathbf{R}$ into polygonal cells so that all points within a single cell ultimately map to the same point. We do not explicitly construct this tessellation but simply discard unnecessary coinciding points. Thus, we obtain the vertices of a dodecagonal tiling, and one can verify that this is the Penrose tiling P3 [35].

We stress that the above algorithm requires the contribution of $\mathbf{r}_0$ to the values of $\phi_k$ to be zero. The coefficient relating $\mathbf{a}_k^{||}$ and $\mathbf{b}_k^{||}$ is determined by the number of plane waves included in Eq. (3); its value is such that if the *round* function in the calculation of $n_k^j$ is replaced by the identity transformation, then the identity $(\mathbf{R}^j)' \equiv \mathbf{R}^j$ holds for any initial point $\mathbf{R}^j$; see more detail in Appendix A.

Secondly, we select those P3 vertices, for which the total mismatches are small and, consequently, the values of $\delta\rho\left(\mathbf{r}_j\right)$ are large. Figure 1(a) shows that the P1 tiling includes those P3 vertices, for which $\delta\rho > 1.6$ ($\Delta < 0.25$). However, some of such vertices with large value of $\delta\rho$ may be absent in the ideal P1 tiling. We emphasize that similarly located additional SUs are observed in real structures usually rationalized on the basis of the P1 tiling, for example in Al-Mn-Si quasicrystal (see Fig. 2 in Ref. [39]).

Nonetheless, the classical P1 tiling can be obtained within the developed theory by introducing the following geometric constraint: from each pair of vertices corresponding to the intense maxima and separated by a short diagonal of a wide rhombus [its length equals $(\sqrt{5}+1)/2+2$ measured in edge lengths of the P3 tiling], it is necessary to select the vertex for which $\delta\rho$ is greater and, accordingly, the total mismatch is smaller. This P1 construction algorithm works when we consider those P3 vertices for which the total mismatch is less than the threshold value lying within the interval $\Delta\subset[0.2,0.25]$. Note that the considered geometrical constraint may be unnecessary or different when describing real structures. For example, instead of geometric constraints, one can introduce an additional interaction energy between SUs, as in Ref. [40], where the authors consider the self-assembly of particles in an external field.

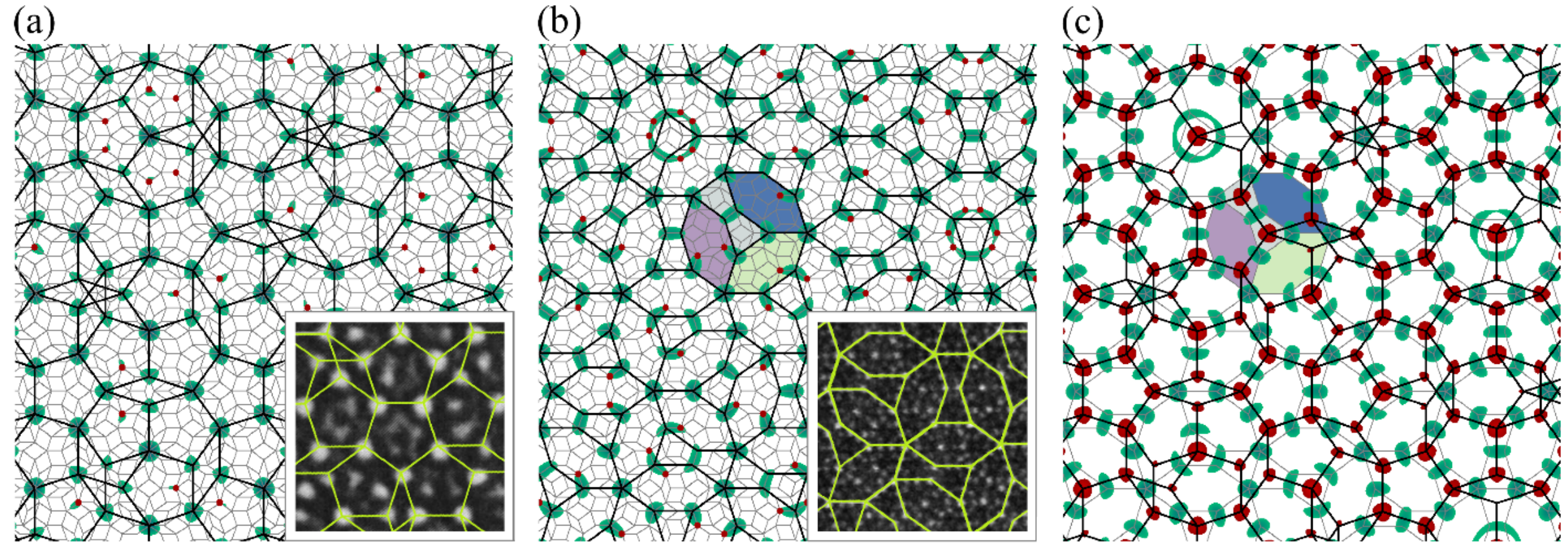

FIG. 1. Decagonal tilings constructed within Landau theory. (a) Penrose P1 tiling with bold edges. It is superimposed on the P3 tiling and has the edges $2\left[\cos\left(\frac{\pi}{10}\right)+\cos\left(\frac{3\pi}{10}\right)\right]\approx 3.078$ times longer. For the shown critical density function, $c_0=0$; similar superimpositions can be obtained for other integer values of $c_0$. P3 vertices with small mismatches but violating the geometrical constraint are marked with small red circles. (b) Tie and Navette tiling superimposed on the P3 tiling and on the decagonal critical density wave with $c_0=1/2$. In both panels (a) and (b), the density functions are shown with a green gradient only within the regions where $\delta\rho(\mathbf{R})>1.6$. The insets show electron microscope images of Al-Mn-Si and Al-Cr-Fe quasicrystals; the images are adopted from Refs. [39,41]. (c) Superimposition of the tilings shown in (a) and (b); the regions where the critical density function for the Tie and Navette tiling is maximal and minimal are shown in green and red, respectively.

If $F_4<0$, the potential minimization results in half-integer values of the phase sum $c_0$. Like in the previous case, the algorithm for finding the maxima of $\delta\rho$ initially leads to the P3 tiling. Intense maxima of the function $\delta\rho>1.6$ can correspond from one to 3 vertices of this tiling. From each pair of these vertices separated by the long diagonal of the narrow rhombus of the P3 tiling, we select the vertex with the larger $\delta\rho$ value. As the result, the Tie and Navette tiling emerges; see Fig. 1(b), where the vertices excluded by the above selection are shown in red. The tiling construction algorithm is applicable if we consider those P3 vertices for which the total mismatch is less than the threshold value lying within the interval $\Delta\subset[0.2,0.36]$.

Note that for a long time, this tiling, obtained back in 1990 [42], was not observed in real systems, and electron microscope images of decagonal Al–Cr–Fe–Si and Al–Cr–Fe quasicrystals, which are similar to the tiling, were first published only in 2020 [41,43]. The Tie and Navette tiling was originally derived as a P1 sublattice [42]. In the proposed approach, the substitution $\Delta\phi\rightarrow\Delta\phi+1/2$ simultaneously leads to a switch between the conditions on $c_0$ and a change in the sign of the density function. Therefore, if we construct and superimpose two such tilings [see Fig. 1(c)], they appear to be dual and the nodes of each of them will coincide with the inter-nodes of the other. Such type superposition of the P1 and Tie and Navette tilings can be found in Ref. [44], but the edge length of the Tie and Navette tiling arising in our theory is smaller.

## III. LANDAU CRYSTALLIZATION THEORY FOR DODECAGONAL TILINGS

In this Section, without reiterating the main ideas of Landau theory, we solely focus on the differences between dodecagonal and decagonal cases. The following subsection considers a pair of dual dodecagonal tilings. The selection between them is governed by the sign of the coefficient associated with the cubic invariant of the nonequilibrium potential. In subsection B within the framework of Landau theory, we obtain an S-T tiling, which serves as the foundational model for describing various structures.

### A. Dual dodecagonal tilings

Unlike the decagonal star of wave vectors, the dodecagonal one includes 12 rays: $\mathbf{b}_k^{||} = b_0 \langle \cos\frac{2k\pi}{12}, \sin\frac{2k\pi}{12} \rangle$, where $b_0$ is their length and $k$=0,1…11. In the quasicrystalline phase, $\delta\rho$ reads:

$$\delta\rho(\mathbf{R}) = \sum_{k=0}^{5} 2\rho_{av} \cos\left(2\pi\left(\mathbf{b}_k^{||}\mathbf{R} + \phi_k\right)\right), \qquad (7)$$

where $\rho_{av} = |\rho_k|$, and $\rho_k$ are the complex amplitudes of the corresponding $\delta\rho$ decomposition.

The second-degree invariant has a regular structure, and the dependence on the phases appears already in the third-degree invariant $I_2 = \rho_0\rho_8\rho_4 + \rho_1\rho_9\rho_5 + \rho_6\rho_2\rho_{10} + \rho_7\rho_3\rho_{11}$ that is proportional to the value of $\cos(2\pi c_1) + \cos(2\pi c_2)$, where

$$c_1 = \phi_0 - \phi_2 + \phi_4 \text{ and } c_2 = \phi_1 - \phi_3 + \phi_5. \qquad (8)$$

The phases can be parameterized as $\phi_k = \mathbf{b}_k^{||}\mathbf{r}_0 + \mathbf{b}_k^{\perp}\mathbf{r}_0^{\perp} + \Delta\phi_k$, where $\mathbf{b}_k^{\perp} = b_0 \langle \cos\frac{5k\pi}{12}, \sin\frac{5k\pi}{12} \rangle$, $\Delta\phi_0 = -\Delta\phi_2 = \Delta\phi_4$, and $\Delta\phi_1 = -\Delta\phi_3 = \Delta\phi_5$. This parameterization takes into account the orthogonality of the subspaces; as in Sec. II, the Landau potential does not depend on either $\mathbf{r}_0$ or $\mathbf{r}_0^{\perp}$. The fourth-degree invariants do not depend on the phases. By minimizing the simplest Landau potential of the 4th degree with respect to the phase variables, we obtain that the negative coefficient at the 3rd degree invariant leads to integer values of expressions (8). Our analysis shows that with continuous variation of $\mathbf{r}_0^{\perp}$ and discrete (integer) variations of $c_1$ and $c_2$, the function $\delta\rho$ is rearranged, but it does not change qualitatively: the same tiling can be superimposed on its maxima.

The tiling with nodes near the intense maxima of Eq. (7) (the phase combinations (8) are integers) is well known [45] and is formed by square, triangle, and shield (S-T-Sh) tiles; see Fig. 2(a). It is derived similarly to the P1 tiling. We project starting points from $\mathbf{R}$ to the 6D space and back with an expression analogous to Eq. (6). The only difference is that in this case, the summation is performed over six terms, and $\mathbf{a}_k^{||} = \frac{3}{b_0}\mathbf{b}_k^{||}$. First, the procedure leads to an analogue of the P3 (AP3) tiling, consisting of $30^0$ rhombic, triangular and square tiles. Next, we select all vertices of the AP3 tiling for which $\delta\rho$ is large and, consequently, the total mismatch is smaller than $\Delta$. Starting from $\Delta = 0.52$ at least one such vertex is located near each maximum where $\delta\rho$>1.6. Then, in the vicinity of each maximum of $\delta\rho$, the vertex with the smallest total mismatch (or largest $\delta\rho$) is selected. Formally, we select the AP3 vertex with the smallest total mismatch from all pairs of vertices separated by a distance less than the edge length of the constructed tiling (which is $\sqrt{2} + \sqrt{2+\sqrt{3}}$ measured in edge lengths of the AP3 tiling). Figure 2(a) shows the excluded vertices as red circles. The proposed algorithm for tiling construction works when the threshold value for the total mismatch is greater than 0.52.

Note that the S-T-Sh tiling can be easily rearranged into a tiling of squares, triangles, and narrow rhombuses [45], by adding one more position inside each shield so that a single $30^0$ rhombus is formed. Out of the three options for adding a position into a shield, the position with the least mismatch is selected. Such positions often appear in areas where $\delta\rho > 1.6$.

A notable aspect of our approach to the S-T-Sh tiling is that inside each regular dodecagon, maxima and minima of the $\delta\rho$ function form two deformed pentagons; see Fig. 2(c). Our comparison of extremum positions of the $\delta\rho$ function and the electron density function of the columnar liquid quasicrystal [36] shows that the maxima of $\delta\rho$ coincide almost perfectly with the minima of the quasicrystal electron density; see this density in Ref. [36], Fig. 4.

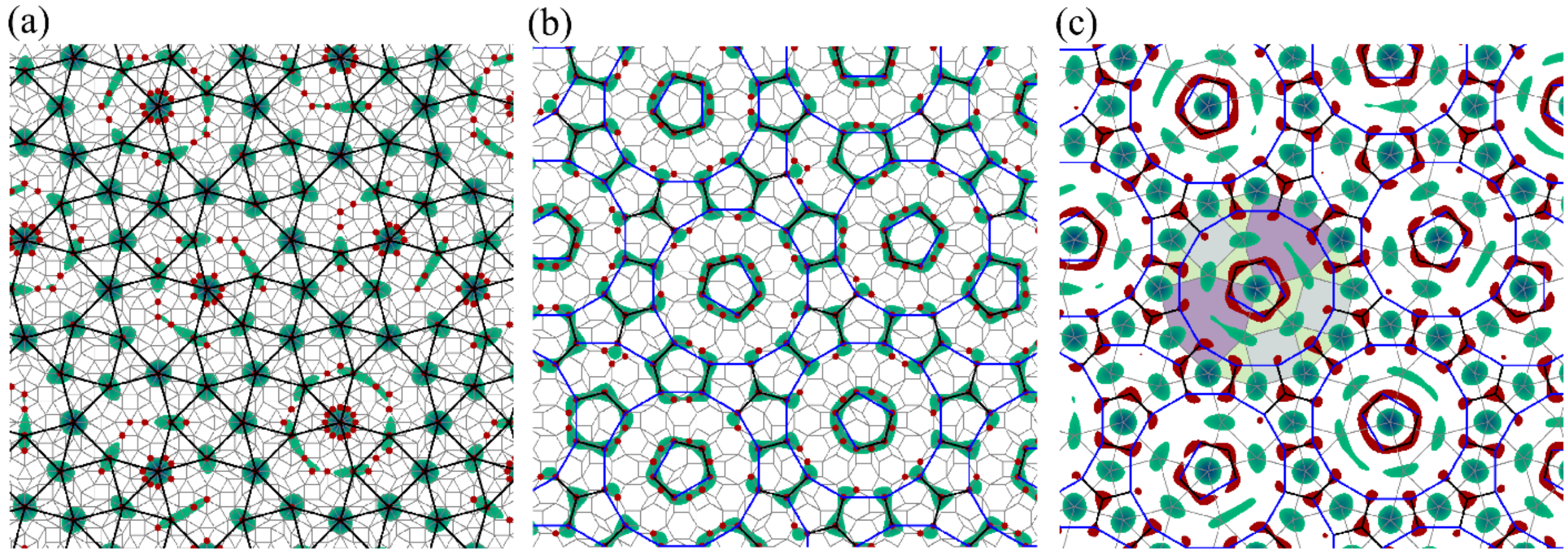

FIG. 2. Dodecagonal tilings and density waves. (a) Tiling composed of squares, triangles, and shields that is superimposed on the analogue of the P3 tiling; in the corresponding dodecagonal critical density function $\delta\rho$, the phase combinations (8) are integers. (b) Tiling composed of dodecagons, their fragments, and deformed pentagons that is superimposed on the second variant of the dodecagonal analogue of the P3 tiling; in this case, the values of combinations (8) are half-integers. The areas in panels (a) and (b), where $\delta\rho > 1.6$, are highlighted in green. (c) Superimposition of the tilings (a) and (b); the areas where $\delta\rho(\mathbf{R})$ for the first tiling is maximal and minimal are shown in green and red, respectively.

The structural model [36] relates the arrangement of minima of the electron density with a dodecagonal tiling of squares, triangles, and narrow rhombuses: the SUs in the form of columns decorate all the vertices of squares and triangles, and are also located in the geometric centers of narrow rhombuses. As the result, there are 5 SUs inside each regular dodecagon around its center. Comparing with our approach, it is easy to see that the location of the SUs in the model [36] corresponds very well to the location of the areas, in which $\delta\rho > 1.6$.

Without entering into a debate with the authors [36], who placed the SUs at the minima of the electron density rather than at its maxima, we pursue our initial goal and construct a tiling dual to the S-T-Sh one. This tiling is energetically favorable when the contribution of the 3rd degree invariant is positive and the values of phase combinations $c_1$ and $c_2$ are half-integer. In Figure 2(b), the tiling is superimposed on the corresponding Landau density function, shown in the regions where $\delta\rho > 1.6$. The algorithm for constructing the tiling is similar to the ones developed above. It is, however, interesting to note that if the values of $c_1$ and $c_2$ are half-integers, then the AP3 tiling is arranged differently and consists of narrow rhombuses, squares, and regular hexagons.

In order to construct the tiling, we first determine those AP3 vertices for which the total mismatch is smaller than the threshold value. When the latter is greater than 0.52, at least one vertex of the AP3 tiling is located near each intense maximum of $\delta\rho$. Then, from each pair of vertices that are closer to each other than the larger diagonal of the narrow rhombus of the AP3 tiling, we select the vertex with the smallest total mismatch. The tiling construction algorithm works when the limit value is greater than 0.52. A characteristic feature of the resulting tiling is the irregular shape of the pentagons: the pentagon edges have two different lengths. As in the decagonal case, a shift of all phases in $\delta\rho$ by 1/2 changes the sign of the critical density wave. Therefore, for a pair of dual tilings, the nodes of one will coincide with the inter-nodes of the other; see Fig. 2(c).

The columnar liquid quasicrystal [36] can transform into a structurally similar periodic phase with a square lattice. In terms of the theory considered, such periodic structures known as approximants emerge due to an appropriate small shift of $\mathbf{b}_k^{||}$ vectors, while the distorted critical density function $\delta\rho'(\mathbf{R})$ still determines the positions of tiling vertices. Assuming that the cubic invariant in the Landau potential is preserved under such shifts, then the deformed vectors $\left(\mathbf{b}_k^{||}\right)' = \mathbf{B}_k$ in the approximants must satisfy the same equations as the original vectors: $\mathbf{B}_0 - \mathbf{B}_2 + \mathbf{B}_4 = 0$, $\mathbf{B}_1 - \mathbf{B}_3 + \mathbf{B}_5 = 0$, and $\mathbf{B}_{k+6} = -\mathbf{B}_k$. Accordingly, the phase combinations $c_1$ and $c_2$ must be either both integer or both half integer. Approximants with the density functions $\delta\rho'(\mathbf{R})$ and $-\delta\rho'(\mathbf{R})$ are dual, and here, we discuss in more detail only the case of integer $c_1$ and $c_2$. The presence of the 4-fold axis removes any arbitrariness except for the choice of a pair of wave vectors $\mathbf{B}_0$ and $\mathbf{B}_2$, and a pair of corresponding phases. Furthermore, since the change in phases should not lead to a shift in the function $\delta\rho'(\mathbf{R})$ as a whole, the following relations must be satisfied: $\phi_0 = -\phi_2 = \phi_4$, and $\phi_1 = -\phi_3 = \phi_5$.

It seems reasonable to place the 4-fold axis at the coordinate origin, which further restricts the possible values of the phases $\phi_i$ to either 0 or 1/2. Since the function $\delta\rho'(\mathbf{R})$ have the translational symmetry of the square lattice, the vectors $\mathbf{B}_k$ must be the vectors of its reciprocal space. For the approximant shown in Fig. 3, $\mathbf{B}_0 = \frac{1}{a} < 4, -1 >$ and $\mathbf{B}_2 = \frac{1}{a} < 3{,}3 >$, where $a$ determines the approximant periodicity. In the case considered, two non-equivalent variants of the function $\delta\rho'(\mathbf{R})$ are possible. In the first one, $\phi_0 = \phi_1 = 0$, and in the second, $\phi_0 = 0$ and $\phi_1 = 1/2$. The approximant shown in Fig. 3 corresponds to the second variant. In its structure, the centers of narrow rhombuses are located on the local 2-fold axes, and the maxima of $\delta\rho'$ perfectly coincide with the locations of columns in the model [36]. Thus, in our opinion, the density wave approach (see Fig. 2 and Fig. 3) provides

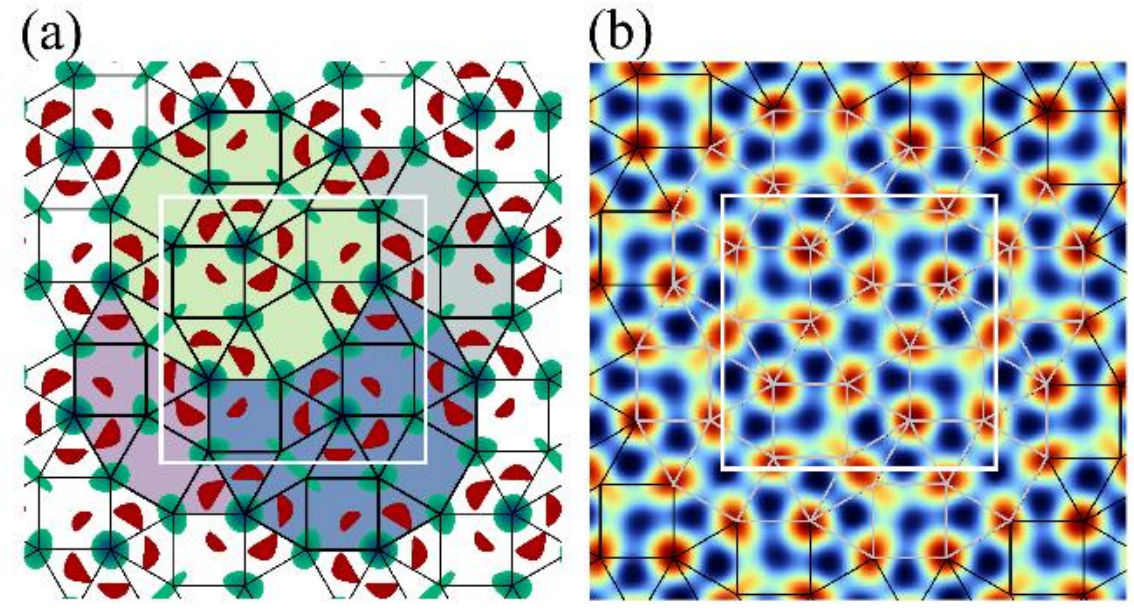

FIG. 3. Approximant of the S-T-Sh dodecagonal tiling [see Fig. 2(a)] with a square lattice. (a) The approximant is superimposed on its density function $\delta\rho'$, which is visualized using the same color code as in Fig. 2(c). (b) The same approximant is superimposed on the electron density function [36], whose value decreases when the color changes from dark orange to dark blue.

a better rationalization of the observed structures than the purely tiling interpretation.

### B. Landau theory and square-triangular tiling

First, let us emphasize that it is impossible to obtain defect-free square-triangular (S-T) tilings by cut-and-project or analogous methods of N-dimensional crystallography; therefore, alternative approaches are required [22, 46-47]. Let us consider how an S-T tiling can be obtained using the approach of critical density waves.

Note that the S-T and S-T-Sh tilings are related because the vertices of both packings are indexed by the same basis vectors. Therefore, as we will demonstrate below, the S-T tiling can be obtained using the critical density wave that we previously derived for the S-T-Sh tiling.

Figure 2(a) shows that the basis translations of the S-T-Sh tiling are rotated by $\pi/12$ relative to the translations of the AP3 tiling, with the edges of the S-T-Sh tiling being $\omega = \sqrt{2} + \sqrt{2+\sqrt{3}}$ times longer. Thus, the basis translations of the S-T and S-T-Sh tilings can be explicitly written as follows

$$\mathbf{a}_k^{\mathrm{ST}} = \frac{3\omega}{b_0}\langle \cos\left(\frac{5k\pi}{12} + \frac{\pi}{12}\right), \sin\left(\frac{5k\pi}{12} + \frac{\pi}{12}\right)\rangle \qquad (9)$$

where k=0,1…11.

In the developed model, we assemble the QC cluster within an expanding circular region by sequentially adding new vertices (new cluster particles) of the corresponding tiling one after another, ensuring that each newly added vertex is separated from the already present ones by one of the translations given by Eq. (9). Also, we assume that particles are added sequentially, within an expanding ring region defined by two radii, $\mathrm{R_{min}}$ and $\mathrm{R_{max}}$ and vacant positions are located at minimum distance $|\mathbf{a}_k^{\mathrm{ST}}|$ from already filled ones. With each particle adding, the radii are recalculated. We define them as the minimum and maximum distances from the origin to vacant positions. The value $\mathrm{R_{max}} - \mathrm{R_{min}} = \Delta\mathrm{R}$ is a model parameter, and at each step we remove from the set of vacant positions all those that are located beyond the circle with the radius $\mathrm{R_{max}}$. At each step, the particle is added provided it has such coordinates $\mathbf{r}_j$ that the value of the function $\delta\rho(\mathbf{r}_j)$ is maximal. If there are several vacant positions that correspond to the same maximum value of the function, then the position to be added is chosen randomly among them. Note that the addition of new particles also maximizes the effective OP (5) as was the case for constructing the tilings discussed above. However, here, the maximization is conditional as it occurs under additional constraints imposed by the model.

If $\Delta\mathrm{R} = 0$, the model leads to the formation of a triangular lattice; if $\mathrm{R_{max}}$ is not limited, the model results in structures with significant unfilled areas. At $\Delta\mathrm{R} \approx 0.2$, an S-T region is formed, in which various defects are formed as it grows; see Fig. 4(a). Recall that various defects are typical of soft matter systems and their stabilization can be attributed to entropic effects [48]. As $\Delta\mathrm{R}$ increases, a tiling containing shields [similar to that shown in Fig. 2(a)] emerges.

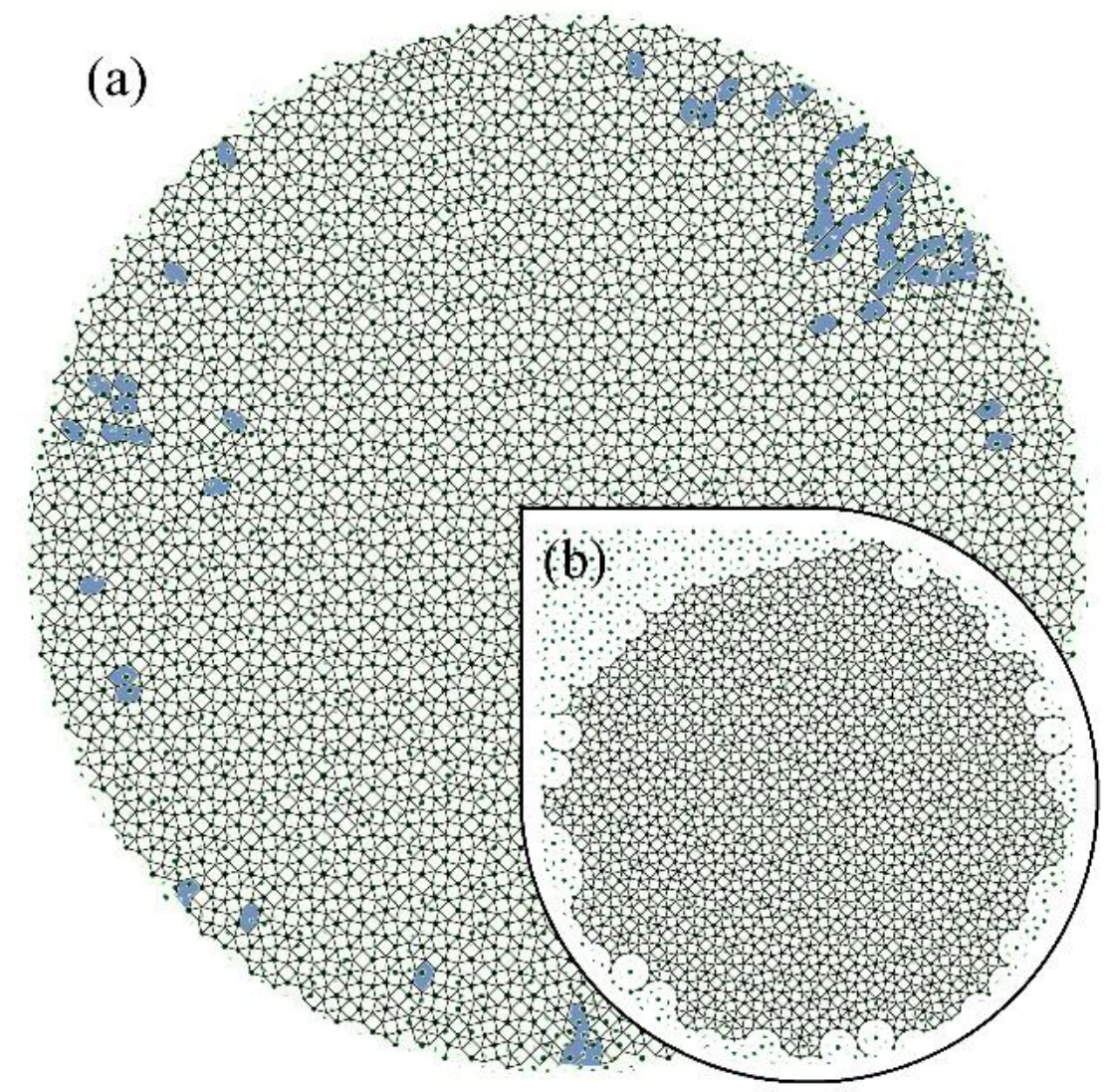

FIG. 4. Square-triangle (S-T) packings obtained within Landau theory. (a) S-T cluster with defects at the perimeter. The completion of tiles was not performed during the cluster assembly; $\Delta\mathrm{R} \approx 0.2$. (b) Defect-free S-T cluster. Square tiles and pairs of triangular ones were completed as soon as possible; $\Delta\mathrm{R} \approx \tau$. Both structures in panels (a) and (b) are superimposed on the critical density wave shown in the areas where $\delta\rho > 1.6$.

One can achieve a higher structural perfection of a growing S-T cluster. To do this, following the ideas [7,24,49], square and triangular tiles should be completed: if, at some stage of growth, three adjacent filled positions turn out to be the vertices of a defective square or a defective pair of triangular tiles then the fourth position is added to eliminate the defect. At $\Delta\mathrm{R} \approx \tau$, where $\tau = \sqrt{3} + 2$ is the self-similarity coefficient of dodecagonal order, the defect-free S-T cluster grows infinitely; see Fig. 4(b). The behavior of the presented model closely resembles that of the S-T cluster growth model [22], which, unlike the proposed one, is based on comparing perpendicular distances for vacant positions in 4D space.

## IV. NONEQUILIBRIUM ASSEMBLY OF RANDOM TILINGS STRUCTURALLY SIMILAR TO P1

An interesting application of the density wave theory lies in the modeling of random tilings structurally similar to ideal ones. As an example, we consider the nonequilibrium assembly of a random P1-type tiling. Like in the previous Section, the process of attaching particles (tiling vertices) to a growing cluster occurs only in those positions that are connected to the nearest occupied ones by at least one translation from a set of 10 orientationally equivalent ones $\pm\mathbf{a}_k^{||}$, where $k$=0,1…4. For simplicity, we chose the normalization of Eq. (6) so that $|\mathbf{a}_k^{||}|$=1. Particle attachment occurs sequentially, and at each step, the number $q$ of vacant positions that can be filled by only one particle is determined first. For the $j$-th position, the probability $p_j$ of its filling is found using the usual Boltzmann distribution

$$p_j = \exp\left(-\frac{E_j - \mu}{T}\right),$$

where $E_j$ is the binding energy, $T$ is the thermal energy, and the chemical potential $\mu$ is determined from the condition $\sum_j p_j = 1$, since exactly one particle is attached. Note that in the limit $T \to 0$, the attached particle has the most favorable (maximally negative) binding energy. In this limit, if we suppose $E_j = -\delta\rho(\mathbf{r}_j)$, the model developed in this section becomes equivalent to the one from the previous section.

Here, we take the thermal energy into account and calculate the binding energy $E_i$ in a more general form. For a particle that can occupy the position with coordinates $\mathbf{r}_j$, the binding energy is given by

$$E_j(\mathbf{r}_j) = \sum_i V(\mathbf{r}_i, \mathbf{r}_j), \qquad (10)$$

where $V(\mathbf{r}_i, \mathbf{r}_j)$ is a pair potential, and the summation is performed over all previously filled positions of the cluster. Considering the pivotal role of critical density waves in the theory of self-assembly, below we propose two pair potentials that are simply related to $\delta\rho$ and lead to the assembly of random P1-type clusters.

Such clusters with few defects can be obtained [see Fig. 5(b)] using the first pair potential, which we refer to as the local wave potential. It is constructed using the critical density wave function $\delta\rho(\mathbf{l})$ [see Fig. 5(a)] as: $V(\mathbf{r}_i, \mathbf{r}_j) = \alpha(\delta\rho(\mathbf{l}) - 0.7)$, where $\mathbf{l} = \mathbf{r}_i - \mathbf{r}_j$, and $\alpha$ is a step function depending on $l_0$: if $l \leq l_0$, then $\alpha = -1$, otherwise $\alpha = 0$. Since the regular pair potential does not depend on the rotation of a pair of particles, the function $\rho(\mathbf{l})$ must be as symmetric as possible, i.e., with $\phi_k = 0$. This condition and the choice of normalization $|\mathbf{a}_k^{||}|$=1 result in the redefinition of $\delta\rho$ as

$$\delta\rho(\mathbf{R}) = \sum_{k=0}^{4} \cos(2\pi \mathbf{b}_k^{c} \mathbf{R}), \qquad (11)$$

where $\mathbf{b}_k^{c} = \frac{4}{5}\left[\cos\left(\frac{\pi}{10}\right) + \cos\left(\frac{3\pi}{10}\right)\right] \langle \cos\left(\frac{2\pi k}{5} + \frac{\pi}{10}\right)$, $\sin\left(\frac{2\pi k}{5} + \frac{\pi}{10}\right) \rangle$; the length of the vectors $\mathbf{b}_k^{c}$ and their rotation (equal to $\pi/10$) are determined from the superposition of the P3 and P1 tilings, as shown in Fig. 1(a).

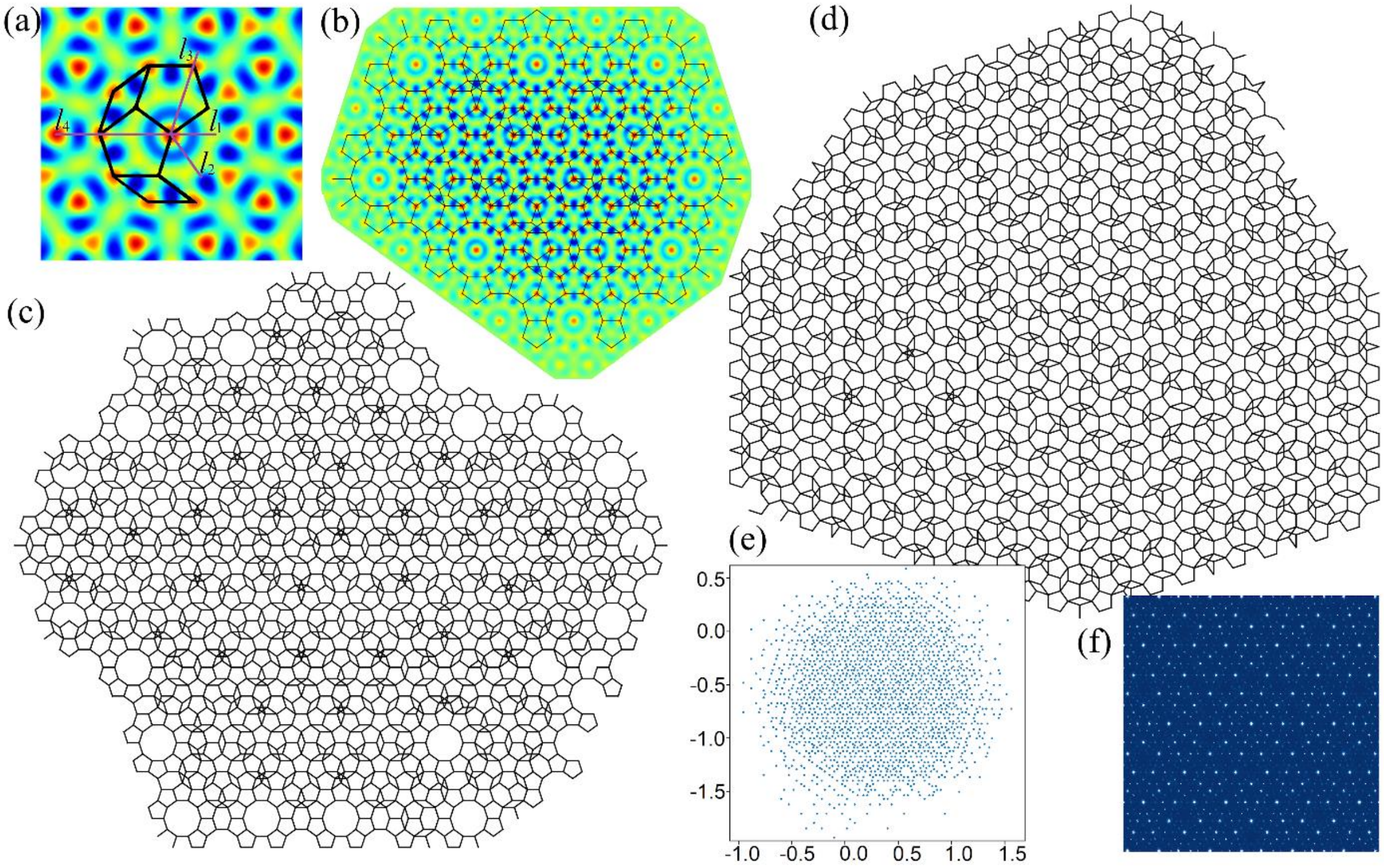

FIG. 5. Nonequilibrium assembly of clusters structurally similar to the Penrose P1 pentagonal tiling. (a) Fragment of a tiling superimposed on the critical density wave with $\phi_k = 0$. The color change from blue to red corresponds to an increase in the function value from -5 to 5. (b) Cluster of 300 particles assembled in the approximation of the local wave potential. The potential created by the cluster is shown with the same color code. (c) Cluster of 2000 particles also assembled in the local wave potential approximation. When assembling the clusters (b-c), $l_0 = 5$ and T = 0.5. (d) More structurally perfect cluster of 2000 particles assembled using the discrete pair potential at T=0.5. (e) Perpendicular coordinates of the cluster (d) positions. (f) Diffraction pattern of the cluster (d).

Figure 5(b) shows a cluster of 300 particles assembled with this potential; the color change from blue to red corresponds to an increase in the absolute (negative) value of the binding energy $E_j$. Figure 5(c) shows a cluster of 2000 particles assembled under the same conditions as the cluster (b). Individual defects are visible inside the cluster (c). The empty decagons around the perimeter of the clusters may be filled as it continues to grow.

Since in the model, particles attach to the cluster only at discrete positions determined by the basis translations $\pm\mathbf{a}_k^{||}$, instead of the continuous local wave potential, one can introduce a simpler discrete pair potential $V(l)$ that is nonzero only for several specific interparticle distances.

Within this approach, clusters with a higher degree of structural perfection can be obtained; see Fig. 5(d). The second developed by us pair potential $\mathrm{V}(l)$ reads: $\mathrm{V}(l_n) = [-1.62;\ V_2;\ V_3;\ -4.41]$, where the four distances $l_n = [1, \sim 1.176, 2, \sim 2.618]$ are the lengths of the parallel projections of vectors <1,0,0,0>, <1,-1,0,0>, <2,0,0,0>, and <1,1,0,-1>. The latter vectors are given in the 4D basis of vectors $\mathbf{a}_i^{||}$, where $i$=0,1,2,3.

The potential is also related to the critical density wave approach: negative (energy-favorable) interactions are calculated as $-\delta\rho(\mathbf{l}_i)$, and points $\mathbf{l}_1$ and $\mathbf{l}_4$ practically coincide with the maxima of $\delta\rho(\mathbf{l})$; see Fig. 5(a). The values of $V_2$ and $V_3$ are chosen sufficiently large; specifically, $\mathrm{V}_2 \geq 30$ and $\mathrm{V}_3 \geq 30$, which eliminates the appearance of particle pairs at distances $\mathbf{l}_2$ and $\mathbf{l}_3$ as well as the corresponding point defects. Points $\mathbf{l}_2$ and $\mathbf{l}_3$ are close to the minima of $\delta\rho(\mathbf{r})$; see Fig. 5(a). Note that, unlike the maxima, the minima of $\delta\rho(\mathbf{r})$ are not indexed by integer indices, so the vectors directed from the origin to the corresponding nearest deep minima are only approximately equal to $\mathbf{l}_2$ and $\mathbf{l}_3$.

The morphological perfection of the obtained clusters can be assessed based on the diffractograms and the dispersion of the perpendicular coordinates of the cluster particles. For clusters obtained using a discrete potential, this dispersion is approximately 10% smaller in diameter; see Fig. 5(e). The diffractograms of the clusters obtained using both types of potentials are similar [see Fig. 5(f)] and demonstrate the presence of dodecagonal symmetry along with long-range order. The diffraction intensity is calculated as $I(\mathbf{q}) = A(\mathbf{q})A^*(\mathbf{q})$, where $\mathbf{q}$ is the reciprocal space vector. The structural amplitude $A(\mathbf{q})$ reads $A(\mathbf{q}) = \sum_n \exp(i\mathbf{q}\mathbf{r}_n)$, where $i = \sqrt{-1}$, and the summation is performed over all particles of the cluster. Let us note that the intensity of the central point of the diffraction pattern is $I(0) = N^2$. To enhance the clarity of Fig. 5(f), we used shades of blue for intensities ranging from 0 to $2 \cdot 10^5$, whereas all regions with higher intensities were colored in plain white.

Thus, we have shown that critical density waves arising in Landau crystallization theory, can be used to construct two pair potentials that are well-suited for modeling the assembly of random tilings with a P1-type structure. A similar methodology can be employed to generate random tilings of other types. For instance, in Ref. [33], a random octagonal tiling derived from the Ammann-Beenker tiling is obtained using analogous approach. According to our yet unpublished results, random S-T and S-T-Sh tilings can also be obtained within the framework of the nonequilibrium assembly.

## V. DISCUSSION AND CONCLUSION

Before discussing the possibility of deriving other tilings within the framework of Landau theory, we note that phase transitions can relatively rarely be controlled by order parameters that span different irreducible representations. For example, the 10-fold P3 tiling (which appears in the paper as a stepping stone in generating decagonal tilings) can be superimposed on the irreducible density wave (11), which is critical for the P1 tiling. However, in such a superposition, not all nodes of the tiling coincide with the maxima of the function; see Fig. 6(a). Nonetheless, the P3 tiling can be superimposed well onto the interference pattern of two irreducible decagonal functions with equal amplitudes; see Fig. 6(b). The critical function (11) for the P1 tiling is taken as the first one of these functions. The wave vectors of the second function belong to the star of vectors $(\mathbf{b}_k^{c})'$, where $(\mathbf{b}_k^{c})' = 2/5\,[1 + 2\cos(\pi/5)] \times \langle \cos(2\pi k/5), \sin(2\pi k/5) \rangle$ and these vectors are expressed in terms of $\boldsymbol{b}_k^{c}$ as simple integer linear combinations. The tiling shown in Fig. 6 has a 10-fold symmetry axis located at the origin, therefore, all phases in both irreducible density functions are zero.

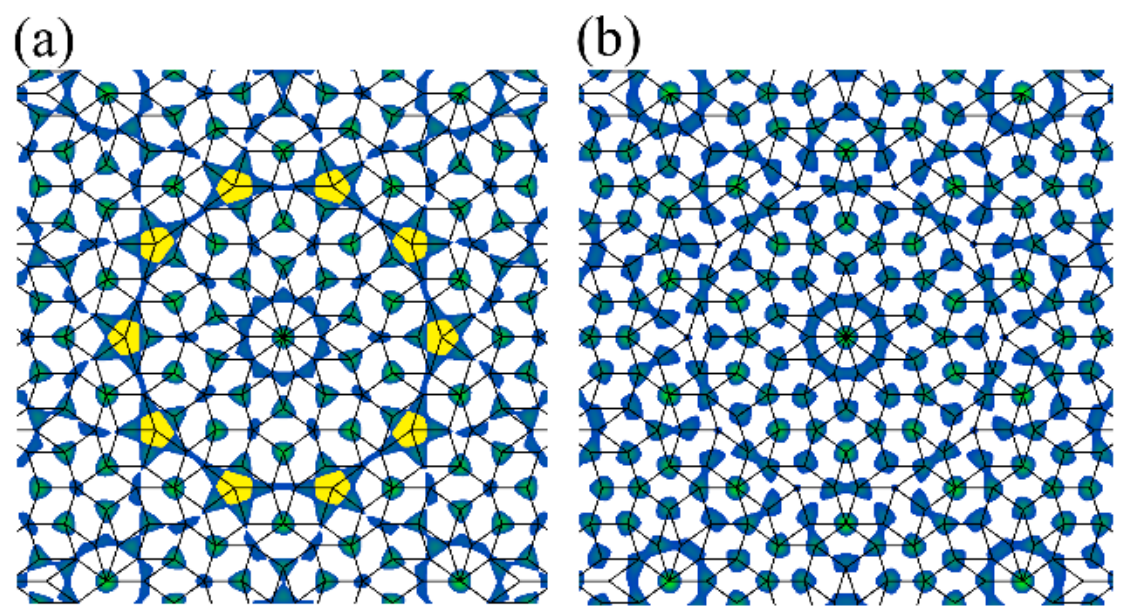

FIG. 6. P3 tiling and density waves. (a) The P3 tiling is superimposed on the critical density function for the P1 tiling. The latter has edges of the same length as those of the P3 tiling. In the regions shown in yellow, the P3 vertices overlap not with the maxima but with the minima of the density function. (b) The P3 tiling is superimposed on the sum of two irreducible decagonal density functions; see the main text.

From the viewpoint of Landau theory, crystallization concurrently controlled by two irreducible representations is less probable. Therefore, the P3 tiling is not observed in real systems; at least, we have not been able to find suitable examples. Nevertheless, the concept of critical density waves can still be applied to model the nonequilibrium assembly of this tiling, but its complexity from the viewpoint of Landau theory leads to a more defective structure and necessitates additional assumptions to model the assembly; see more details in Appendix B.

Next, we find it interesting to qualitatively compare the order in the real systems with decagonal and dodecagonal symmetries. In the first (decagonal) case, these are mainly 2- or 3-component metal alloys with a fairly perfect quasicrystalline order. Electron microscope images of many such structures reveal second-generation tiling [7], and the corresponding diffraction patterns show reflections with large index values. In contrast, most dodecagonal structures consist of equivalent structural units (SUs) and belong to soft matter systems. Although the presence of long-range quasicrystalline order in soft matter systems is also confirmed by the corresponding diffraction patterns, it is less perfect. Nevertheless, the observation of diffraction patterns with clearly resolved multiple peaks indicates the possibility of applying the phenomenological density wave approach in both decagonal and dodecagonal cases. In this context, we note that in recent years dodecagonal order has been discovered in a number of different substances that are neither soft matter systems nor metallic alloys[17-21,50].

As a limitation of the theory developed in Sections II and III, we note that it always leads to idealized packings characterized by infinite coherence length and the absence of phason fluctuations. The limitation is eliminated by the approach of nonequilibrium assembly that is well suited for modeling structures with random phason fluctuations. The approach is relevant to Landau theory since it employs potentials obtained from critical density waves. To our knowledge, in the framework of classical molecular dynamics (MD), it is impossible to obtain either the decagonal P1 tiling or the Ammann-Binker one, which is typical of the octagonal QCs. In addition, MD always leads to very defective structures. For example, in packings generated by the recent MD model of the decagonal Tübingen triangle tiling [51], there are not only random phason fluctuations but also a significant number of various defective tiles, which are absent in the ideal tiling. In contrast, by employing the nonequilibrium assembly model, the number of defective tiles can be minimized, and random phason fluctuations can be controlled by temperature.

The classical MD approach is more suitable for modeling dodecagonal soft matter structures, where the order is the least perfect compared to other quasicrystalline systems. In combination with the two-minimum Lennard-Jones-Gauss (LJG) potential, the approach reproduces defective dodecagonal packings similar to the S-T-Sh tiling [52]. According to our results (which will be published elsewhere), such tilings can also be reproduced within the framework of the nonequilibrium assembly. For this purpose, it is sufficient to use the simplest discrete potential that accounts for only two energetically favorable interactions at distances $l_1 = 1$ and $l_2 \approx 1.897$ at which the LJG potential has two minima. As in the case of the nonequilibrium assembly of the P1 tiling, the binding energies are chosen to be proportional to the amplitude of the critical density wave. We emphasize that even in the dodecagonal case, the nonequilibrium assembly model is more fruitful than the MD approach. By completing the tiles (see Sec. III.B) one can generate random packings formed by just two types of tiles (square and triangular ones) as is often observed in soft matter systems [48,50]. By adding terms to the discrete potential, the phason disorder can be decreased, while varying the temperature can conversely increase it.

Comparing the relative complexity of the assembly of decagonal and dodecagonal tilings, we note that in the latter case, the simplest Landau potential is of a lower 4$^{th}$ degree, and, accordingly, the real (non-phenomenological) interactions leading to dodecagonal structures can be simpler. This is consistent with the fact that dodecagonal packings in contrast to decagonal ones are often assembled from one sort of SUs, which obviously simplifies the nature of interactions in relevant dodecagonal systems. In this context, we note that the simplest potential necessary for nonequilibrium assembly of the P1 tiling involves interactions beyond the second coordination sphere and, therefore, it is more complex than the simplest potential necessary for nonequilibrium assembly of the square-triangular tiling.

In discussing the limitations of the nonequilibrium assembly model, let us note that although the Tie and Navette tiling, as well as the dodecagonal tiling with deformed pentagons described in this paper, are simple from the viewpoint of Landau theory, we have not yet succeeded in obtaining simple pair potentials leading to these packings. The main reason for this is that, when constructing pair potentials based on density waves, one must assume that $\phi_k = 0$, which ultimately leads to the assembly of tilings that are dual to the intended ones. However, random Tie and Navette tiling and random tiling with deformed pentagons can be obtained from their counterparts if one takes advantage of the superimpositions of dual tilings found in this paper. We stress that random tilings consisting of the same tiles as the initial perfect ones may be of interest for a variety of applications, such as studying localized states in photonic and phononic quasicrystals [53] or even analyzing possible Hamiltonian cycles in them [54].

The density wave approach can also be useful for rationalizing periodic approximant structures. As we have shown in Sec. III, the analysis of approximants within the density waves approach can be more informative than conventional multidimensional crystallography. In this context, we note that short-period dodecagonal and octagonal approximants can also serve as the basis of structural models of numerous spherical protein shells [22,33].

In conclusion, we hope that the Landau theory and the approach of density waves will continue to be in demand for investigating various quasicrystal structures These theoretical frameworks provide valuable insights into understanding complex ordering phenomena beyond traditional multidimensional crystallography, enabling the design and discovery of new materials with unique physical properties.

## ACKNOWLEDGMENTS

S. R., A.R., and O. K. acknowledge financial support from the Russian Science Foundation, Grant No. 22-12-00105-П. The authors are grateful to D. Chalin for fruitful discussions.

## APPENDIX A: P3 AND ANALOGOUS TILINGS OBTAINED FROM N-DIMENSIONAL SIMPLE CUBIC LATTICES

A by-product of the developed theory is a method for generating the P3 tiling and its analogs with 2n-fold symmetry. Note that P3-like tilings can be derived as the projections onto the 2D plane $E^{||}$ of certain nodes of the corresponding nD cubic lattices. These nodes are located in the strip, which is parallel to $E^{||}$ and has a width of one cell of the nD lattice. We present the same concept in a slightly different but equivalent form, compatible with the proposed algorithm for finding the most intense maxima of the interference patterns.

For convenience, we assume that the edges of the nD lattice have unit length and surround the lattice nodes by cubic Voronoi cells [55]. If the hyperplane $E^{||}$ intersects a Voronoi cell, then its center (which is the node of nD lattice) is projected onto $E^{||}$. In the nD space, this hyperplane can be parameterized by a 2D vector $\mathbf{R}$ as $\{\mathbf{R}\mathbf{b}_i^{||} + \phi_i\}$, where $\mathbf{b}_i^{||}$ are the parallel projections of the reciprocal basis vectors of the cubic lattice and $\phi_i$ are the components of a constant nD shift vector. The latter is orthogonal to the hyperplane: namely, it is orthogonal to the vector with components $\alpha\mathbf{b}_i^{||}$, where $\alpha$ is an arbitrary 2D constant. Then the coordinates of the center of the dissected $j^{\text{th}}$ Voronoi cells read $\{n_i^j\} = \{round(\mathbf{R}^j\mathbf{b}_i^{||} + \phi_i)\}$, where the function $round$ is rounding to the nearest integer and $\mathbf{R}^j$ is an arbitrary starting point from the intersection area. Using the center coordinates, we obtain its projection onto $E^{||}$ as:

$$\mathbf{r}_j = \sum_{i=0}^{N-1} n_i^j \mathbf{a}_i^{||}, \qquad (12)$$

where translations $\mathbf{a}_i^{||} = \langle \cos\frac{2\pi i}{N}, \sin\frac{2\pi i}{N} \rangle$ are the tiling edges with unit length: $|\mathbf{a}_i^{||}| = 1$. In this case

$$\mathbf{b}_i^{||} = 2\mathbf{a}_i^{||}/N, \qquad (13)$$

where $N$ is the dimensionality of the space. The relation (13) guarantees that when the $round$ function is replaced by the value of its argument, any point $\mathbf{R}$ from the space $E^{||}$, after mapping to the nD space and applying Eq. (12), returns to its original position.

Note that the absence of starting points in the intersection area results in the loss of the corresponding tiling vertex. It can occur if the Voronoi cell has a small intersection area with the hyperplane $E^{||}$. Since the centers of such cells are located far from the hyperplane, they are characterized by a large total mismatch and do not correspond to intense maxima of $\delta\rho$ function. Also, we note that for the tilings satisfying Landau theory, the subspace $E^{||}$ is located in a specific way relative to the 5D or 6D spaces: it passes either through the nodes or through the inter-nodes of the corresponding nD cubic lattices.

Let us consider in a little more detail the relation between $\mathbf{b}_i^{||}$ and $\mathbf{a}_i^{||}$. Note that expressions (12) are the first two rows of the projection matrix from the nD space to the subspace $E^{||}$. The other rows of the matrix project the nD space onto the orthogonal complement of $E^{||}$. Since all rows of the matrix are constructed to be mutually orthogonal, their next pairs can be chosen as $\mathbf{a}_i^{\perp} = \langle \cos\frac{2\pi mi}{N}, \sin\frac{2\pi mi}{N} \rangle$, where $m$ is not a divisor of $N$. For the 5D and 6D spaces, there is only one such pair of rows; as the dimensionality of the space increases, the number of such pairs increases and they correspond to different vector representations of the star symmetry group. Also, the projection matrix may include rows reflecting the presence of integer dependencies between $\mathbf{a}_i^{||}$ in $E^{||}$ space. For example, if we use the numeration of vectors that corresponds to (12), then the last row in the 5D projection matrix has to be chosen proportional to <1,-1,1,-1,1>.

By inverting the projection matrix, one can obtain the reciprocal space vectors $\mathbf{b}_i$ that satisfy the usual relations $\mathbf{b}_i \mathbf{a}_j = \delta_{ij}$. Further analysis results in Eq. (13).

## APPENDIX B: NONEQUILIBRIUM ASSEMBLY OF RANDOM TILINGS WITH P3-LIKE STRUCTURE

The model can lead to the assembly of random defective tilings, structurally similar to the Penrose tiling P3. An appropriate local wave potential $V(\mathbf{r}_i, \mathbf{r}_j)$ can be expressed in terms of the function $\rho(\mathbf{l})$, which is the sum of two irreducible density waves [see $\rho(\mathbf{l})$ in Fig. 7(a)]. Namely, $V(\mathbf{r}_i, \mathbf{r}_j) = \alpha(\rho(\mathbf{l}) + 1)$, where $\mathbf{l} = \mathbf{r}_i - \mathbf{r}_j$ and $\alpha$ is the step function defined in the main text. The function $\rho(\mathbf{l})$ is taken to be maximally symmetric, i.e., with $\phi_i = 0$. Panel (b) shows a cluster of 2000 particles assembled within this potential at T=0.5 and $l_0 = 5$. Under these parameters, the least defective clusters are assembled; see Fig. 7(b).

Nonetheless, slightly more morphologically perfect clusters can be obtained [see Fig. 7(c)] by using a more complex discrete pair potential $\mathrm{V}(l)$. It is nonzero for six interparticle distances $l_{\mathrm{n}}$ only: $\mathrm{V}\,(l_n) = [V_1;\ V_2 - 1;\ V_4; -1;\ -1.8]$, where $l_n = [0.618, 0.727, 1, 1.328,\ 1.902,\ 3.078]$ are the lengths of the parallel projections of the vectors <1,0,0,1>, <2,0,2,1>, <1,0,0,0>, <1,1,-1,1>, <1,0,0,-1> and <1,1,-1,-1>, respectively. In this potential, the negative interactions are approximately equal to $-\rho(\mathbf{l}_{\mathrm{I}})$, and the points $\mathbf{l}_3$, $\mathbf{l}_5$, and $\mathbf{l}_6$ practically coincide with the maxima of $\rho(\mathbf{l})$. The values of $V_1, V_2$ and $V_4$ are chosen to be sufficiently large (greater than 5) to prevent the formation of particle pairs at the corresponding distances. The minima of $\rho(\mathbf{r})$ are not indexed by integer indices, so the vectors pointing from the origin to the nearest minima are only approximately equal to $\mathbf{l}_1$, $\mathbf{l}_2$ and $\mathbf{l}_4$. If the assembly is carried out in an unconstrained area, the model produces a highly defective structure formed by differently overlapping decagons. By imposing a restriction on the growth region, it is possible to obtain clusters with packing similar to the P3 tiling; see Fig. 7(c).

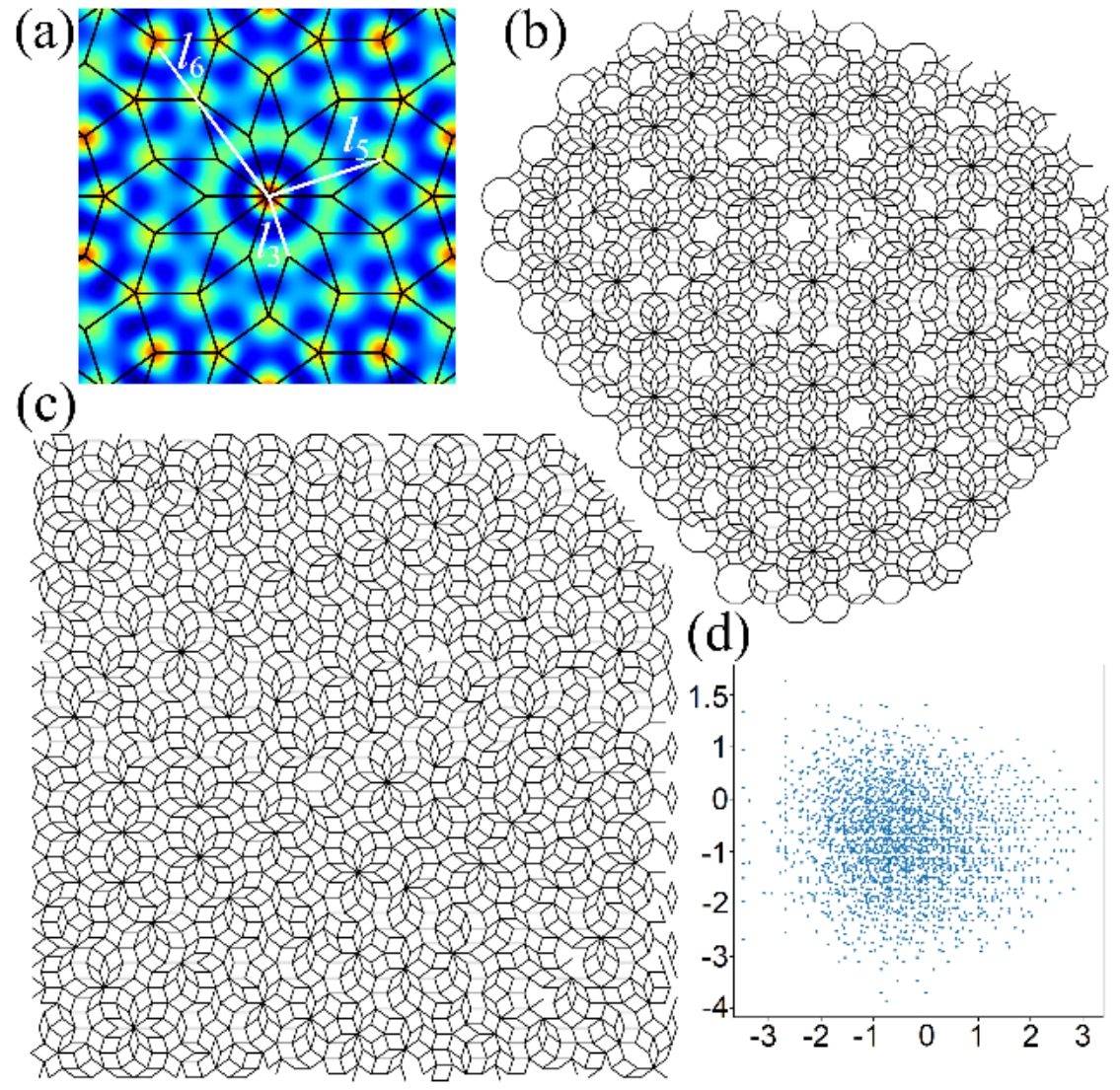

FIG. 7. Nonequilibrium assembly of clusters structurally similar to the Penrose P3 packing. (a) Fragment of the packing superimposed on the interference pattern of two critical density waves with zero phases at the origin. The color change from blue to red corresponds to the function increase from -10 to 10. (b) Cluster of 2000 particles assembled using the local wave potential: $\alpha_0 = 5$ and T=0.5. (c) More structurally perfect cluster of 2000 particles assembled using the discrete pair potential at the same temperature. (d) Perpendicular coordinates of the cluster (c) positions.